# Single-molecule force spectroscopy reveals structural differences of heparan sulfate chains during binding to vitronectin

Katarzyna Herman[1], Joanna Zemła[2], Arkadiusz Ptak[1] and Małgorzata Lekka[2*]

[1]*Institute of Physics, Faculty of Materials Engineering and Technical Physics, Poznan University of Technology, Piotrowo 3, PL-60965 Poznan, Poland*
[2]*Department of Biophysical Microstructures, Institute of Nuclear Physics, Polish Academy of Sciences PL-31342 Kraków, Poland*

*\*Malgorzata.Lekka@ifj.edu.pl*

The syndecans represent an ongoing research field focused on their regulatory roles in normal and pathological conditions. Syndecan's role in cancer progression becomes well-documented, implicating their importance in diagnosis and even proposing various cancer potential treatments. Thus, the characterization of the unbinding properties at the single molecules level will appeal to their use as targets for therapeutics. In our study, syndecan-1 and syndecan-4 were measured during the interaction with the vitronectin HEP II binding site. Our findings show that syndecans are calcium ion-dependent molecules that reveal distinct, unbinding properties indicating the alterations in heparin sulfate chain structure, possibly in the chain sequence or sulfation pattern. In that way, we suppose that HS chain affinity to ECM proteins may govern cancer invasion by altering syndecan ability to interact with cancer-related receptors present in the tumor microenvironment, thereby promoting the activation of various signaling cascades regulating tumor cell behavior.

**PACS:** 87.64.Dz, 82.39.-k, 87.80.Nj, 87.10.-e,

## I. INTRODUCTION

Syndecan (SDC) family is composed of two large (syndecan-1 and syndecan-3) and two small (syndecan-2 and syndecan-4) members [1]. These molecules, located in the plasma membrane, belong to transmembrane heparan sulfate type I proteoglycans that interact with numerous ligands such as growth factors [2], enzymes [3], and extracellular matrix (ECM) proteins such as fibronectin, vitronectin, collagens, or laminins [4]. Various research has already reported a significance of syndecans expression in normal and disease-related functioning of cells [5–7]. Syndecan-1 (SDC-1) is primarily expressed in mesenchymal and epithelial cells, while syndecan-4 (SDC-4) is typical for a wide variety of cell types [8]. SDC-1 plays an essential role in regulating inflammation or chemokine gradient formation for trans-endothelial and trans-epithelial migrations of neutrophils [9,10]. SDC-4 is mainly associated with cellular adhesion and migration, where a synergistic link between integrins and syndecans has been postulated and partially demonstrated [11–13]. SDC-1 regulates αvβ3 or αvβ5 binding that affects, for example, the angiogenesis in cancer cells [14,15]. SDC-4 is frequently associated with β1 and β3 integrins in focal adhesions [16,17].

Syndecans are linear structures consisting of extracellular, transmembrane, and cytoplasmic parts [1,18]. The cytoplasmic part, specific for each syndecan, comprises two highly conserved domains flanking the variable region. Its typical roles are linking syndecans to actin filaments, binding to PDZ proteins, or promoting syndecan-specific signaling. The transmembrane domain participates in syndecan oligomerization. A conserved phenylalanine residue regulates the interactions among syndecans by promoting the formation of either homodimers or heterodimers [19]. The extracellular domain (ectodomain) contains cleavage sites for various metalloproteinases, releasing syndecans' ectodomains in the ECM. The interactions of syndecans with integrins involves either an NXIP motif (SDC-4, [20]) or a synstatin part (SDC-1, [21]). Ectodomains bear covalently attached glycosaminoglycans (GAGs) that, depending on the member type of the syndecan family, are composed of chondroitin (CS) and/or heparan (HS) sulfate chains. SDC-1 contains both CS and HS, while SDC-4 possesses only three HS chains (a single model has been built only for a syndecan ectodomain with three HS chains, but no experimental data confirm this structure, [22]). The binding site of CS chains to syndecan-1 is membrane-proximal, while the HS chains are located at sites at the N-end of the SCD-1 protein core. In SCD-4, predominantly HS chains are bound at the N-end, usually attached to serine/glycine sites on the syndecan core [1,23].

Unfortunately, despite the gathered extensive knowledge about syndecans' involvement in integrin-mediated adhesion, the full details describing their participation are still lacking. Besides, the lack of complete structural data showing the entire ectodomains makes it challenging to predict how GAGs influence syndecans' interaction with ECM proteins. Moreover, little is known about the dynamics of the GAGs conformation changes accompanying the binding to ECM proteins, thus limiting the understanding of the syndecans' role in various processes, such as signal transduction or integrin-mediated adhesion. We have recently shown that the expression of SDC-4 is significantly larger than the expression of SDC-1 in bladder cancer cells. This supports the overall findings stating that the increase or decrease in the abundance of a specific member of the syndecan family has severe consequences affecting the invasion and progression of various cancers [5,24].

In our study, we employed an atomic force microscope (AFM) for single-molecule force spectroscopy (SMFS) [25,26] to study the binding of syndecans (SDC-1 and SCD-4) to ECM proteins (vitronectin, VN). The syndecans bind to the heparan-binding site (HEP II, 345-378 AA) located between two hemopexin-like domains at the C-end of the vitronectin [27]. The later studies narrowed the binding sequence to 12 amino acids (Lys(Asp)347 to Gly358) located within the heparin's initially defined binding domain [27–29]. All syndecans bind to the same HEP II binding site. Therefore, we ask ourselves what the degree of specificity in such a recognition process is. Our study hypothesizes that unbinding of single SDC-VN complexes processes are similar for similar binding configurations, regardless of the syndecan type. Therefore, we expect to obtain similar thermodynamic and kinetic parameters from the force spectroscopy analysis for SDC-1 and SDC-4 complexes. To probe the unbinding properties of an individual SDC/VN complex, SDC-1 and SDC-4 were immobilized on a freshly cleaved mica surface while vitronectin was attached to the surface of silicon nitride cantilevers.

## II. EXPERIMENTS: MATERIALS AND METHODS

### A. Recombinant proteins

Vitronectin and syndecans were recombinant proteins purchased from R&D Systems. Syndecan-4 (SDC-4, $M_w$ = 24 kDa; SDS-PAGE, reducing conditions) consists of a signaling sequence (18 AA) attached to an extracellular domain (127 AA), a transmembrane region (25 AA), and a cytoplasmic tail (28 AA). Syndecan-1 (SDC-1, $M_w$ = 85 kDa, SDS reducing conditions) consists of a signaling sequence (22 AA) attached to an extracellular domain (232 AA), a transmembrane region (21 AA), and a cytoplasmic tail (35 AA). An extracellular matrix protein – vitronectin (VN, Asp20-Leu478, 70−80 kDa SDS-PAGE, reducing conditions) consisted of a signaling peptide (19 AA) and a protein (459 AA). The amino terminated end (130 AA) contains multiple binding sites, including an RGD sequence that binds to integrins, while the carboxyl-terminated end contains a heparin (HEP II) binding site. Its predominant structure is a monomer.

### B. Monoclonal antibodies

The specific interactions involving syndecans were probed with monoclonal antibodies. They were MabSDC1 (A-6, Santa Cruz Biotechnology, Inc.), recognizing amino acids sequence from 82–256 AA, an extracellular domain in SDC-1 of human origin, and MabSDC4 (5G9, Santa Cruz Biotechnology, Inc.), recognizing a sequence from 93–121 AA, an extracellular domain in SDC-4 of human origin. Solutions containing monoclonal antibodies were prepared by using PBS buffer.

### C. Mica surface modification

Syndecans (SCD-1 and SDC-4) were immobilized on an atomically flat, freshly cleaved muscovite mica surface (about 0.25 $cm^2$). Mica surface was silanized with 3-aminopropyl-triethoxysilane (1 mL, APTES, Signa-Aldrich placed in a Petri dish) for 1.5 in a desiccator. Afterward, the silanized mica surface was activated with glutaraldehyde (2.5%) aqueous solution for 20 minutes. After glutaraldehyde activation, the mica surface was gently washed with phosphate-buffered saline (PBS, Signa-Aldrich). Syndecans were dissolved in the PBS buffer at the concentration of 0.2 μg/mL (molar concentrations were of 2.4 nM and 8.3 for SDC-1 and SDC-4, respectively). A drop of such a solution was placed on the mica surface for 30 minutes. Afterward, samples were thoroughly rinsed with PBS buffer to remove the excess of the unbound material.

### D. Cantilever functionalization.

To measure the unbinding force of SDC-1/VN or SDC-4/VN complexes, the surface of silicon nitride cantilevers (PNP-TR, NanoWorld) was coated with vitronectin (0.2 μg/ml dissolved in PBS buffer, the molar concentrations of 8.0 nM and 13.6 nM, respectively) for 25 minutes. Cantilevers were pre-treated using an analogous procedure as for the mica surface described above (1.5h of silanization, 2.5% glutaraldehyde activation, PBS rinsing). After incubation with vitronectin, cantilevers were washed three times with the PBS buffer (2 minutes) and kept in the PBS buffer prior to measurements.

### E. Single-molecule force spectroscopy.

Measurements of the unbinding force were conducted using AFM head Force Robot (Bruker-JPK) working in automatized force spectroscopy mode (maximum Z-range of 10 μm). The cantilevers' spring contact (PNP-TR, nominal value of 0.03 N/m, NanoWorld) was determined using thermal fluctuation methods.30 All measurements were conducted in 50 mM Tris (Sigma-Aldrich) supplemented with 1 mM concentrations of $Ca^{2+}$ and $Mg^{2+}$ ions. The use of Force Robot head enabled to carry out measurements in one run (4 – 5 days without a break, for setpoint set to 500 pN) using one cantilever in a broad range of retraction velocities, from 0.1 µm/s to 19.0 µm/s (10 different velocities were chosen). The loading rate, calculated as a product of the effective spring constant (it was obtained from the slope of the force-displacement curve at the rupture event) and the retraction velocity, corresponded to a range from 540 pN/s to 380000 pN/s. The force curve acquisition process was carried out in cycles. Each cycle accounted for a 64 × 64 mesh of points localized on 400 µm² surface area, which delivered 4096 single force curves. The cycles were carried out pseudo-randomly versus the retraction velocities to avoid the effect of tip wear in time. Each measurement cycle was acquired using the same cantilever. Measurements were repeated twice. All the force curves were analyzed with JPK SPM data processing software. There was no need to use a polymer linker because of the large molecules forming the studied complexes.

### F. Assuring specificity of syndecan binding.

The unbinding measurements were carried out in different environments to demonstrate the specificity of the SCD-1/VN and SDC-4/VN interaction, namely, the PBS buffer and the PBS buffer supplemented with 10 mM EDTA (Sigma-Aldrich). We also applied monoclonal antibodies (MAbs) against syndecans. In this case, the mica surface was functionalized with the solution containing a mixture of syndecan: Mabs (in 1:1 ratio, concentrations of 0.2 μg/mL; molar concentrations were 2.4 nM and 8.3 nM for MabSDC1 and MabsSDC4, respectively) to block all syndecan molecules. In such a case, we obtained samples in which ~29% and ~50% molecules were blocked for SDC-1/VN and SCD-4/VN complexes (Table 1).

TABLE 1. Theoretical accessibility of binding sites in the inhibition of the syndecans (0.2 μg/mL) by mixing them with the corresponding monoclonal antibodies (0.2 μg/mL).

| Surface | molar concentrations | syndecan status |
|---|---|---|
| SDC-1 on mica | 8.03 nM | 100% of free SDC-1 are accessible |
| SDC-4 on mica | 13.61 nM | 100% of free SDC-4 are accessible |
| SCD-1/MabSDC1 (1:1) on mica | 8.03 /2.35 nM | 71% of free SDC-1 are accessible |
| SDC-4/MabsSDC4 (1:1) on mica | 16.61 /8.33 nM | 50% of free SDC-4 are accessible |

## III. DYNAMIC FORCE SPECTROSCOPY

The DFS spectra serve as a basis for the thermodynamic and kinetic characterization of the unbinding process, delivering the information on the interacting molecules' energy landscape. A few theoretical models describe a stochastic character of the escape process from a potential well such as Bell-Evans (BE, considers only the position of the energy barrier treated as a point, [30–31]), Dudko-Hummer-Szabo (DHS, considers realistic shape of the energy barrier [32]), and Friddle-Noy-De Yoreo (FNDY, considers reversible bond formation, [33]). They can be applied to estimate the kinetic parameters of the unbinding process between single molecules. The BE model assumes that the applied external force lowers the height of the energy barrier. Evans and Ritchie have derived from the Bell formula the most probable unbinding force [31]:

$$F = \frac{k_B \cdot T}{x_\beta} \cdot ln\left(\frac{r_f}{F_0 \cdot k_{off}^0}\right) \quad (1)$$

where $k_B$ is the Boltzmann constant, $T$ is the absolute temperature; $r_f$ is the loading rate, $x_\beta$ is the distance between the maximum of an activation barrier and the bound state minimum (on the free-energy landscape), and $k^0_{off}$ is the force-free dissociation rate. By applying a linear regression to DFS data, one can easily extract the values of $x_\beta$ and $k^0_{off}$ that is why the BE model has been widely applied to study specific interactions between molecules, including the receptor-ligand complexes [32–38]. There are two main limitations of this model: it ignores the rebinding processes and reduces all the information about the energy landscape profile (the interaction potential) to a single parameter – $x_\beta$. The DHS model goes beyond the BE by specifying the energy landscape profile and applying Kramer's diffusion theory [32]. In contrast to the BE model, it enables to extract $\Delta G_\beta$ additionally to $x_\beta$ and $k^0_{off}$, from the following expression for the most probable unbinding force:

$$F = \frac{\Delta G_\beta}{\nu \cdot x_\beta} \cdot \left\{ 1 - \left[ \frac{k_B \cdot T}{\Delta G_\beta} \cdot ln\left(\frac{k_{off}^0 \cdot k_B \cdot T \cdot e^{\left(\frac{\Delta G_\beta}{x_\beta}\right)}}{x_\beta \cdot r_f}\right) \right]^\nu \right\} \quad (2)$$

where $\Delta G_\beta$ is the free energy of activation in the absence of external forces and the parameter ν is related to the shape of the free-energy potential. Its value equaled to 2/3, although specific for the linear-cubic, is appropriate for all smooth free-energy surfaces. The model has already been applied to analyze the interaction occurring in ligand-receptor binding [39,40] or self-assembled monolayers [41–43]. Both models,

BE and DHS, ignore the possibility of reversible bond formation during force-induced unbinding experiments. This term has been introduced in the FNDY model that considers rebinding's contribution in the unbinding process described by Bell's formula [33]. The defined equilibrium force ($F_{eq}$) describes the force at which the unbinding passes from the near-equilibrium to the kinetic regime. It is related to the force at which the rates of dissociation and association of molecular complexes are in equilibrium:

$$F_{eq} = \sqrt{2 \cdot k_{eff} \cdot \Delta G_{unb}} \quad (3)$$

Here, $k_{eff}$ is the effective spring constant that considers the cantilever's stiffness and the molecular complex, and $\Delta G_{unb}$ is the free energy of the unbound state relative to the bound one. The values of $F_{eq}$ together with $x_\beta$ and the dissociation rate at the equilibrium force – $k_{off@Feq}$ describe the kinetics and strength of an intermolecular bond. The following equation approximates the most probable unbinding force:

$$F \cong F_{eq} + \frac{k_B \cdot T}{x_\beta} \cdot ln\left(1 + e^{-\gamma} \cdot \frac{r_f}{\frac{k_B \cdot T}{x_\beta} \cdot k_{off@Feq}}\right) \quad (4)$$

where $\gamma = 0.577$ is the Euler−Mascheroni constant. The main model application is the unfolding process of titin and the $\beta$–amyloid dimers' rupture [40,44],

## IV. RESULTS AND DISSCUSION

### A. Unbinding of single syndecan/VN complexes.

To probe the unbinding properties of individual SDCs/VN complexes (Fig. 1), SDC-1 and SDC-4 were immobilized on a freshly cleaved mica surface at the concentration of 0.2 µg/mL.

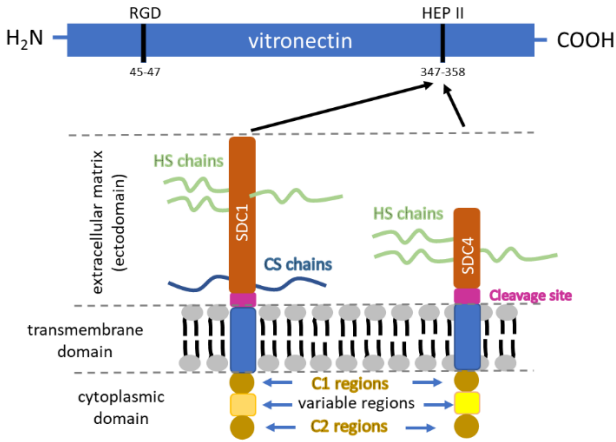

FIG. 1. (online color) Structure of syndecan-1 and syndecan-4 and the interaction with the HEP II binding site located at the C-end of vitronectin.

Vitronectin was attached to the surface of silicon nitride cantilevers (0.2 µg/mL). Despite structural differences, syndecan interaction involves HS chains covalently attached at the N-end in the syndecan core proteins. Thus, intuitively, such a binding seems to proceed similarly. Force curves were collected during the unbinding of the VN-modified AFM probe from the SDCs coated mica surface (Inset in Fig. 2a, which simultaneously presents an exemplary overlay of 20 force curves recorded for SCD-1/VN complex).

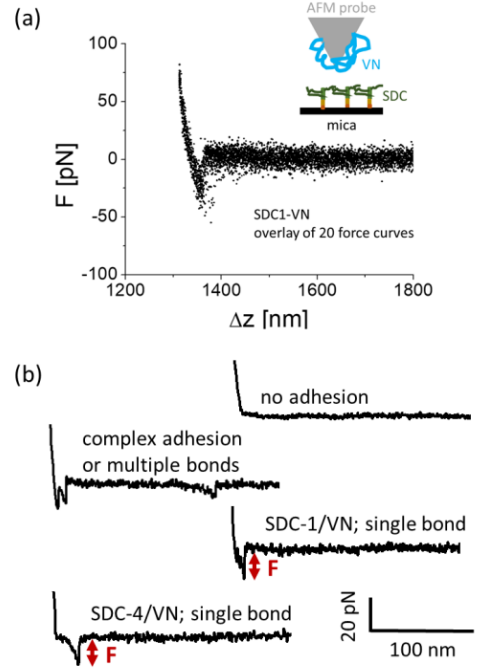

FIG. 2. (online color) (a) An exemplary overlay of 20 force curves recorded for SDC-1/VN complex is presented. (b) Force curves showing no adhesion or complex adhesion, including multiple unbinding events, that were excluded from the analysis. Force curves showing the single-unbinding events characteristic for the rupture of SCD-1/VN and SDC-4/VN complexes.

An area of 400 µm$^2$ was probed by setting a grid of 64 pixels × 64 pixels that resulting in the acquisition of 4096 force curves. Measurements were carried out in 50 mM TRIS supplemented with 1 mM concentrations of Ca$^{2+}$ and Mg$^{2+}$. An individual force curve represents the dependence between the cantilever's vertical deflection (related to the unbinding force) and the tip-sample distance. Force curves (Fig. 2b), observed for SDC/VN interaction, were classified into the groups showing: no adhesion / no specific interaction, a complex adhesion involving the multiple unbinding events, and single unbinding events attributed to a specific interaction of either SDC-1 or SDC-4 to vitronectin. We considered only single molecule unbinding in the analysis, which accounted for about ~8% of all recorded force curves.

### B. Specificity of the syndecan/VN interaction.

The interaction was blocked in various ways and quantified in terms of the unbinding probability to demonstrate the specificity of SDC-1/VN and SDC-4/VN unbinding (Fig. 3). This is the semi-quantitative measure of how many specific unbinding events are observed at given experimental conditions, including retraction velocity. It is usually determined as a ratio between the number of specific unbinding events to the total number of force curves recorded.

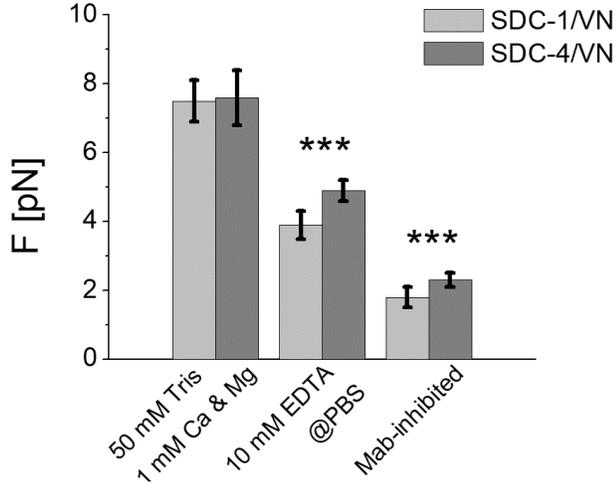

FIG. 3. The unbinding probability showing a decrease upon blocked SDCs/VN interaction (data are mean ± standard deviation (s.d.), *** $p < 0.0001$).

According to previous works, the unbinding probability of about 30% ensures an 83% chance that the measurement reflects a single receptor-ligand unbinding [34]. In the condition of fully accessible binding sites, the probability of the unbinding events for the SDC-1/VN and SDC-4/VN was 7.5 ± 0.6% and 7.6 ± 0.8%, respectively. Such a low value assures that observed unbinding stems from a rupture of a single molecular complex [45,46]. Having in mind the previously reported results [39], the unbinding measurements were conducted in 10 mM EDTA added to PBS. The results show a statistically significant drop in the number of the unbinding events to 3.9 ± 0.4% and 4.9 ± 0.3%, correspondingly. In the experiments when monoclonal antibodies were used to block the SDC and VN interaction, syndecans were mixed with the corresponding monoclonal antibodies at a 1:1 volumetric ratio, followed by the mixture deposition on the mica surface. The SMFS measurements showed a drop in the number of unbinding events to the level of 1.8 ± 0.2% (SCD-1/VN) and 2.3 ± 0.1% (SDC-4/VN), thus confirming that the selected force curves represent the specific recognition occurring between the studied molecules. A summary of the unbinding probability is included in Table 2.

TABLE 2. The unbinding probability for syndecans–VN complexes, measured at various conditions: 50 mM Tris supplemented with 1 mM CaCl2, 1 mM MgCl2; PBS; 10 mM EDTA in PBS buffer, and in conditions of Mab-inhibited interaction. Unbinding probability is expressed as a mean ± standard deviation (s.d.).

| Conditions | SDC-1/VN | SDC-4/VN |
|---|---|---|
| 50 mM TRIS supplemented with 1 mM CaCl2, 1 mM MgCl2 | 7.5% ± 0.6% | 7.6% ± 0.8% |
| 10 mM EDTA in PBS buffer | 3.9% ± 0.4% | 4.9% ± 0.3% |
| Mab-inhibited SDCs-VN interaction | 1.8% ± 0.3% | 2.3% ± 0.2% |

A similar level of the unbinding probability suggests that in the syndecan/vitronectin interactions, all syndecans bind to heparan (HEP II) binding site independently of their structure. However, in addition to results confirming the specificity of the interaction, we can also state that divalent ions are needed for the binding of syndecans to vitronectin. The $Ca^{2+}$ and $Mg^{2+}$ ions probably maintain HS chains' conformation since they are covalently attached to the syndecan N-end that protrudes significantly beyond the cell membrane. However, the more considerable drop in the unbinding probability obtained for SDC-1/VN complex may also indicate differences in HS structures among the studied syndecans. These results agree with already reported data showing that divalent ions like $Mg^{2+}$, together with heparan sulfate chains, could maintain appropriate CS and HS chains' conformation, thereby increasing their accessibility [47].

### C. Comparing unbinding force for SDC-1/VN and SD-4/VN complexes.

All forces related to a single unbinding were gathered into histograms, formed separately for each retraction velocity. Figs. 4ab present exemplary histograms prepared for three distinct retraction speeds, i.e., 0.6 µm/s, 6 µm/s, and 19 µm/s created for both types of molecular complexes. As it is typical for this type of measurement [25], the center and the unbinding event distribution width increase with the retraction velocity. The histograms were fitted with a function describing a lognormal distribution of the events. We have decided to use the mode values in further analysis to be compatible with our previous works [5,39].

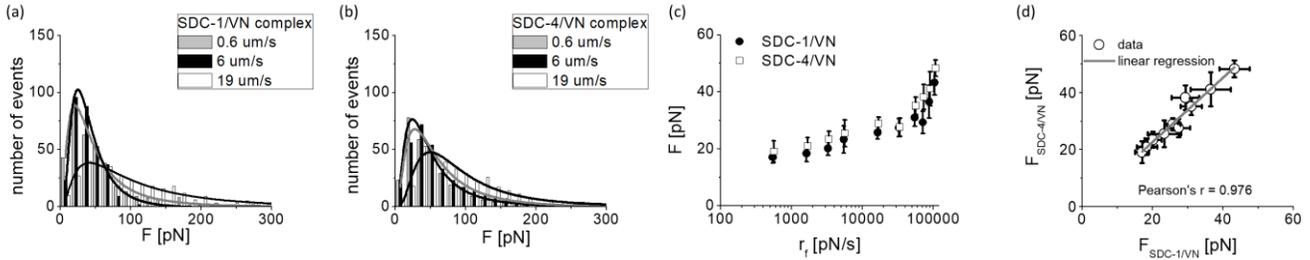

FIG 3. (a,b) Exemplary distributions of the unbinding events obtained for the unbinding of SDC-4/VN and SDC-1/VN ruptured at the retraction velocity of 0.6 µm/s, 6 µm/s, and 19 µm/s. Lines represent lognormal fits. (c) The unbinding force is plotted against the loading rate on a scale logarithmic (data represents mode and s.d.) (d) A correlation between the unbinding force for SDC-1/VN and SCD-4/VN complexes.

Knowing that a simple SMFS measurement of force curve does not deliver sufficient information on the nature of the intermolecular interaction, we applied dynamic force spectroscopy (DFS, [45]) to measure the most probable unbinding force plotted as a function of the loading rate (describing how fast the applied force changes in time during molecular complex unbinding). In our experiments, retraction velocity varied from 0.1 µm/s to 19.0 µm/s what corresponded to loading rates ranging from 540 pN/s to 380 000 pN/s. Then, we compared the obtained dynamic force spectra for both complexes (Fig. 4c). The unbinding force for a single SCD-1/VN complex is smaller than the unbinding force for SCD-4/VN within the whole range of the applied loading rates. The proportionality quantify by linear regression shows a strong correlation between the corresponding unbinding forces ($F_{SDC-4/VN} = 1.13 \cdot F_{SDC-1/VN}$; Pearson's r = 0.976; Fig. 4d). We explain this relation by the participation of only HS chains in the binding with ECM proteins and attribute the observed small changes to structural differences in HS chains interacting with VN. CS chains in syndecan-1 are located closer to the cell membrane (Fig. 1). Thus, steric hindrance may prevent CS chains from reaching the HEP II binding site. Altogether, these results may indicate a conservative-like character of the binding site preserved for all syndecans in the interaction with ECM proteins.

### D. Kinetic and thermodynamic parameters of SCDs/VN complexes.

In our study, the DFS spectra obtained for the unbinding of single SDC-1/VN complexes show two regions with shallow and steep slopes (Figs. 4c, 5). The nonlinear character of the relation between the most probable unbinding force and the loading rate logarithm is similar to already reported data obtained for other biological complexes [25,36,48]. The nonlinear character of the $F$–$ln(r_f)$ relation can be interpreted differently. It may denote a transition through an inner energy barrier that starts to dominate at high loading rates (BE,[46,48]), or it can be explained by an intrinsic feature of a single barrier, namely its smooth shape and hence the force-induced shortening of the distance to the transition state (DHS, [32,49]). Therefore, to get deeper insights into syndecan's unbinding process from vitronectin BE and DHS models were fitted to DFS spectra (Fig. 5). The fits of BE and DHS models to the whole loading rate range resulted in $R^2adj < 0.9$. This value strongly suggests that the nonlinearity in the $F$–$ln(r_f)$ relation originates due to two outer and inner energy barriers.

For the low loading rate range, from 540 pN/s to 34000 pN/s (region I), the rupture forces increase almost linearly, indicating the presence of one single dominant barrier. At higher loading rates, from 34000 pN/s to 110000 pN/s (region II), a steeper slope points to the presence of an inner energy barrier. According to the BE model (and the DHS), two distinct slopes in the force spectrum correspond to the two kinetically distinct energy barriers. The linear region with shallow slope corresponds to the outer energy barrier ($k_BT/x_\beta$ = 4.2 pN·nm 2.8 ± 0.3 pN; the goodness of the BE fit $R^2adj$ = 0.9554, Fig. 3a). The linear regression fitted for higher loading rates delivers the slope of 37.8 ± 1.1 pN ($R^2adj$ = 0.9986), which translates into the kinetic parameters describing the passing over the inner energy barrier. The inner energy barrier only appears when the outer one is significantly lower by force rupturing the molecular complex. Each energy barrier is characterized by its position in relation to the energy minimum and the dissociation rate constant. These parameters for SDC-1/VN complex obtained with the BE model were $x_\beta$ = 1.41 ± 0.12 nm, $k^0_{off}$ = 0.69 ± 0.42 s$^{-1}$ (the outer energy barrier dominating within the low loading rate regime) and $x_\beta$ = 0.111 ± 0.003 nm, $k^0_{off}$ = 872 ± 4 s$^{-1}$ (the inner energy barrier dominating at larger loading rates). The BE model cannot be applied to determine the energy barrier's height because all the information about the energy landscape shape is included in one parameter $x_\beta$. However, it is possible to calculate the height difference by applying the equation comparing dissociation rate constants $|\Delta G_{\beta,inner} - \Delta G_{\beta,outer}| = - k_BT \cdot ln(k^{0,inner}_{off}/k^{0,outer}_{off})$. The obtained difference shows that the outer activation barrier is ~7.14 $k_BT$ larger than the inner one. Thus, the force inducing the unbinding of SDC-1/VN complex must lower this value's outer energy barrier.

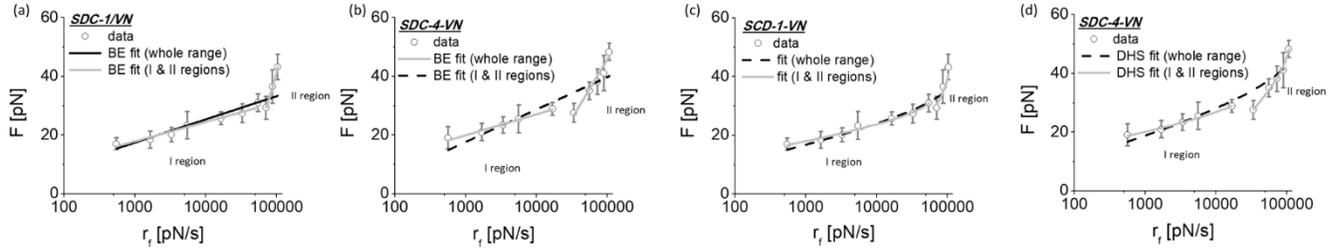

FIG. 5. (Online color) The dynamic force spectrum obtained for the unbinding of SDC–VN complexes. Data (points) were fitted with both BE (a,b) and DHS (c,d) models. Dashed lines are the fits of BE and DHS models to the whole range of the loading rates, i.e., from 540 pN/s to 100000 pN/s. The solid lines show the fits performed for low (from ~ 540 pN/s to 34000 pN/s; region I) and high (34000 to 100000 pN/s; region II) loading rates. The error bars represent standard deviations.

The rupture of the SDC-4 /VN complex seems to proceed along the same pathways manifested in similar DFS spectra. Analogous BE-based analysis of the unbinding of the SCD-4 complex delivers a set of the kinetic parameters:

(i) $x_\beta = 1.31 \pm 0.10$ nm and $k^0_{off} = 0.51 \pm 0.26$ s$^{-1}$ (the outer energy barrier, $R^2$adj = 0.9789), and

(ii) $x_\beta = 0.232 \pm 0.0025$ nm and $k^0_{off} = 418 \pm 56$ s$^{-1}$ (the inner energy barrier, $R^2$adj = 0.9568).

The BE-derived energy height difference is ~6.71 $k_BT$, thus, relatively 0.5 $k_BT$ lower than for the SDC-1/VN complex. To validate the calculations of energy difference between two barriers, simultaneously eliminating the effect of the force-induce shortening of the outer energy barrier's position, the DHS model was fitted to the data for both complexes (Figs. 5cd). The fits of the DHS model to a whole range of the loading rates were characterized by $R^2$adj < 0.9 what supported the hypothesis that the nonlinearity in the $F$–$ln(r_f)$ relation originates due to the presence of two outer and inner energy barriers. Thus, analogously as for the BE model, the DHS one was fitted to the same regions in the DFS spectra resulting in a set of the kinetic parameters. The advantage of this model is the direct determination of the energy barrier heights.

Kinetic parameters, obtained from the DHS model, derive the energy barrier's height for both studied complexes. Their values were similar, i.e., 14.6 ± 1.0 $k_BT$ and 14.2 ± 2.0 $k_BT$ for SCD-1/VN and SDC-4/VN complexes, respectively. The heights of the corresponding inner energy barriers were 2.95 ± 0.47 $k_BT$ and 4.18 ± 0.23 $k_BT$. The difference between the energy barriers for SCD-4/VN is lower than for SDC-1/VN complexes (10.0 $k_BT$ versus 11.6 $k_BT$). In parallel, the lifetimes characterizing the crossing over the energy barriers for SDC-4/VN are higher, too. These results suggest that the force-induced unbinding of syndecan-1 from the HEP II binding site proceeds faster than the SDC-4/VN complex. The thermodynamic and kinetic parameters describing the force-induced unbinding of SDCs-VN complexes are summarizing in Table 3.

TABLE 3. Thermodynamic and kinetic parameters describing the force-induced unbinding of two SDCs-VN complexes.

| Model | Loading rate range | $x_\beta$ [nm] | k0off [s-1] | ΔGβ [kBT] | R2adj |
|---|---|---|---|---|---|
| | | SDC-1 / VN | | | |
| BE | 540 pN/s – 34000 pN/s | 1.41 ± 0.12 | 0.69 ± 0.42 | – | 0.9554 |
| | 34000 pN/s – 110000 pN/s | 0.111 ± 0.003 | 872 ± 4 | – | 0.9986 |
| DHS | 540 pN/s – 34000 pN/s | 2.86 ± 0.16 | 0.018 ± 0.011 | 14.6 ± 1.0 | 0.9815 |
| | 34000 pN/s – 110000 pN/s | 0.439 ± 0.001 | 589 ± 156 | 3.0 ± 0.5 | 0.8598 |
| | | SDC-4 / VN | | | |
| BE | 540 pN/s – 34000 pN/s | 1.31 ± 0.10 | 0.51 ± 0.26 | – | 0.9789 |
| | 34000 pN/s – 110000 pN/s | 0.23 ± 0.03 | 418 ± 56 | – | 0.9568 |
| DHS | 540 pN/s – 34000 pN/s | 2.65 ± 0.81 | 0.013 ± 0.035 | 14.2 ± 2.0 | 0.9868 |
| | 34000 pN/s – 110000 pN/s | 0.53 ± 0.01 | 213 ± 19 | 4.2 ± 0.2 | 0.9900 |

### E. Rebinding in during the unbinding of syndecans/VN complexes.

To account for the rebinding, the FNDY model [33] was applied. This theoretical model considers *(i)* a near-equilibrium regime between the unbinding and rebinding of individual complex. The most probable unbinding force is not dependent on the loading rate, and *(ii)* a kinetic regime where the unbinding force strongly depends on the loading rate. The rebinding affects the unbinding when force curves are collected at low loading rates (Fig. 6).

The FNDY model divides the force spectrum into an equilibrium regime and a kinetic regime. In the former, the rebinding is possible (consequently $x_\beta$ and $k^0_{off}$ deviate from those obtained from the BE model for low loading rates), while in the latter is negligible due to high rupture speed.

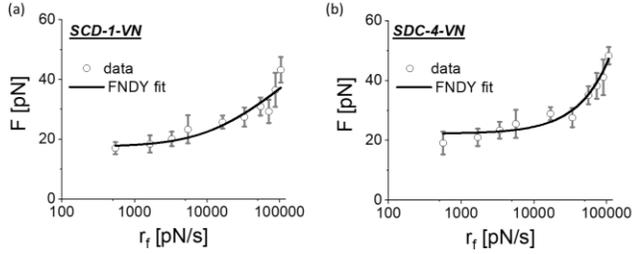

FIG. 6. Analysis of potential rebinding phenomenon based on the FNDY model. Data (points) were fitted with the FNDY model applied to the whole range of the loading rates (540 pN/s to 100 000 pN/s. The error bars represent standard deviations.

TABLE 4. Kinetics parameters for SDCs-VN complexes obtained based on FNDY model.

|  | SDC-1/VN | SDC-4/VN |
| --- | --- | --- |
| $x_\beta$ [nm] | 0.44 ± 0.22 | (9.7 ± 1.8)·1015 |
| $k^0_{off@F_{eq}}$ [s$^{-1}$] | 843 ± 545 | 2397 ± 178 |
| $\Delta G$ [$k_B T$] | 2.10 ± 0.40 | 3.80 ± 0.56 |
| $F_{eq}$ [pN] | 17.5 ± 1.4 | 22.2 ± 1.3 |
| $R^2_{adj}$ | 0.8851 | 0.9374 |

The FNDY model fitted to the data matches DFS spectra for both studied complexes ($R^2adj$ = 0.8851 and $R^2adj$ = 0.9374, Table 4). The transition from equilibrium to the kinetic regime is defined by the $F_{eq}$ being 17.5 ± 1.4 pN and 22.2 ± 1.3 pN for SCD-1/VN and SDC-4/VN. In our experiments, such force values are recorded at the loading rates of 540 pN/s and 1700 pN/s. Thus, the results for the two first loading rates can be affected by the rebinding process. However, in the case of SDC-4/VN complex, the FNDY fit (for the whole loading rate range) gives an unrealistically small values of $x_\beta$ suggesting that the rebinding is insignificant, and the outer barrier is responsible for the $F$–$\ln(r_f)$ relation in the low loading rate range.

## V. CONCLUSIONS

The interaction of syndecans with ECM lies based on mechanotransduction governing and regulating cell fate [50]. Specifically, they are among the essential molecules for microbial infections and cancer development did not mention their accompany in the integrin-mediated adhesion [13,51–53]. Here, we postulated that syndecan binding to HEP II binding sites involves only HS chains linked to the core syndecan protein's outermost part. The studied syndecans (SDC-1 and SDC-4) have appeared to be the Ca ions dependent molecules. Finally, we ask ourselves how kinetic and thermodynamic parameters are related to the structure and role of the syndecan-1 and syndecan-4 binding to the HEP II site in the vitronectin. From available structural data, at first sight, it seems that the interaction occurring between syndecans and vitronectin involves mostly HS chains as they are the most protruding part to the outside of the cell [1,22,23,50].

In our study, syndecan-1 and syndecan-4 interact with the same HEP II binding site of the vitronectin. Therefore, we expected that the binding/unbinding would proceed similarly, regardless of the syndecan type. The syndecan interaction with vitronectin seems to rely mainly on HS chains being their most distant component, while the influence of the CS chains present in SDC-1 seems to be limited due to their membrane-proximal localization. By considering the difference between the dissociation rates and values describing the rupture of SDC-1/VN and SDC-4/VN complexes (Table 3), our results indicate the existence of structural alterations in the HS-related binding site (HEP II), possibly in chain sequence or sulfation pattern [54].

In that way, HS chains' affinity to ECM proteins may govern cancer invasion by altering syndecan ability to interact with cancer-related receptors present in the tumor microenvironment, thereby promoting the activation of various signaling cascades regulating tumor cell behavior.

## ACKNOWLEDGEMENTS

This work was supported by the National Science Centre (Poland) project no UMO-2014/15/B/ST4/04737 (ML). KH and AP acknowledge the Ministry of Science and Higher Education in Poland for financial support. The JPK purchase has been realized under the project cofounded by the Małopolska Regional Operational Program, Measure 5.1 – Krakow Metropolitan Area as an important hub of the European Research Area for 2007-2013, project no MRPO.05.01.00-12-013/15.